\documentclass[aps,showpacs,12pt,groupedaddress,superscriptaddress,amsmath,amssymb]{revtex4-1}

\usepackage[    bookmarks,
                 bookmarksopen = true,
                 bookmarksnumbered = true,
                 linktocpage,
                 colorlinks = true,
                 linkcolor = blue,
                 urlcolor  = blue,
                 citecolor = blue,
                 anchorcolor = green,
                 hyperindex = true,
                 hyperfigures]
                 {hyperref}

\usepackage{multirow}
\usepackage{graphicx,xcolor}

\newcommand{\Msun}{M_{\odot}}

\begin{document}

%% \preprint{TKYNT-XXXX}

\title{Mapping neutron star data to the equation of state using the
  deep neural network}

\author{Yuki~Fujimoto}
\email{fujimoto@nt.phys.s.u-tokyo.ac.jp}
\affiliation{Department of Physics, The University of Tokyo, 7-3-1 Hongo, Bunkyo-ku, Tokyo 113-0033, Japan}

\author{Kenji~Fukushima}
\email{fuku@nt.phys.s.u-tokyo.ac.jp}
\affiliation{Department of Physics, The University of Tokyo, 7-3-1 Hongo, Bunkyo-ku, Tokyo 113-0033, Japan}

\author{Koichi~Murase}
\email{murase@sophia.ac.jp}
\affiliation{Department of Physics, Sophia University, 7-1 Kioi-cho, Chiyoda-ku, Tokyo 102-8554, Japan}

\begin{abstract}
  The densest state of matter in the universe is uniquely realized
  inside central cores of the neutron star.  While first-principles
  evaluation of the equation of state of such matter remains as one of the
  longstanding problems in nuclear theory, evaluation in light of
  neutron star phenomenology is feasible.  Here we show results from
  a novel theoretical technique to utilize deep neural network with supervised
  learning.  We input up-to-date observational data from neutron star
  X-ray radiations into the trained neural network and estimate a
  relation between the pressure and the mass density.  Our results are
  consistent with extrapolation from the conventional nuclear models
  and the experimental bound on the tidal deformability inferred from
  gravitational wave observation.
\end{abstract}

%\pacs{07.05.Mh}
% 07.05.Mh	Neural networks, fuzzy logic, artificial intelligence
\maketitle

%%%%%%%%%%%%%%%%%%%%%%%%%%%%%%%%%%%%%%%%%%%%%%%%%%
\section{Introduction}
%%%%%%%%%%%%%%%%%%%%%%%%%%%%%%%%%%%%%%%%%%%%%%%%%%

Neutron stars provide us with a natural laboratory to
study the densest state of matter in our universe (see
Refs.~\cite{Haensel:2007yy, Lattimer:2012nd, Ozel:2016oaf,
  Baym:2017whm, Blaschke:2018mqw} for recent reviews).  The essential
ingredient for neutron star structures is the equation of state (EoS)
$p=p(\rho)$, i.e., a relation between the pressure $p$ and the mass
density $\rho$.  It is a longstanding challenge to evaluate the EoS
from the first-principles theory.

In the cores of neutron stars the baryon density may reach $\gtrsim 5
\rho_0$, where $\rho_{0}$ is the normal nuclear density $\rho_{0}
\simeq 2.7\times 10^{17}$~kg/m$^{3}$.  At such high density properly
dealing with quantum chromodynamics (QCD) is indispensable.
Symmetries of QCD imply a speculative duality at high density between
hadronic and quark states, called quark--hadron
continuity~\cite{Schafer:1998ef}.
The duality at high density
has been confirmed also in a particular limit of large colors of
quarks, and the dual state was named quarkyonic
matter~\cite{McLerran:2007qj}.  The EoS construction founded on
quarkyonic matter has been proposed~\cite{Fukushima:2015bda,
  McLerran:2018hbz}, which is consonant with the phenomenological EoS
constructions~\cite{Masuda:2012kf, Alvarez-Castillo:2013spa,
  Baym:2017whm,Oter:2019kig}.

Although QCD is the established theory, the first-principles
calculations of the EoS have serious problems.  Among various
theoretical approaches the most powerful method is the lattice-QCD
simulation; however, the standard Monte-Carlo algorithm breaks down at
finite density, dubbed the sign problem (see Ref.~\cite{Aarts:2015tyj}
for a review).
The perturbative QCD (pQCD) calculation is also
feasible~\cite{Kurkela:2009gj}, but it is valid only at asymptotically
high density.

%%%%%%%%%% FIGURE %%%%%%%%%%
\begin{figure}
\centering 
\includegraphics[width=0.65\textwidth]{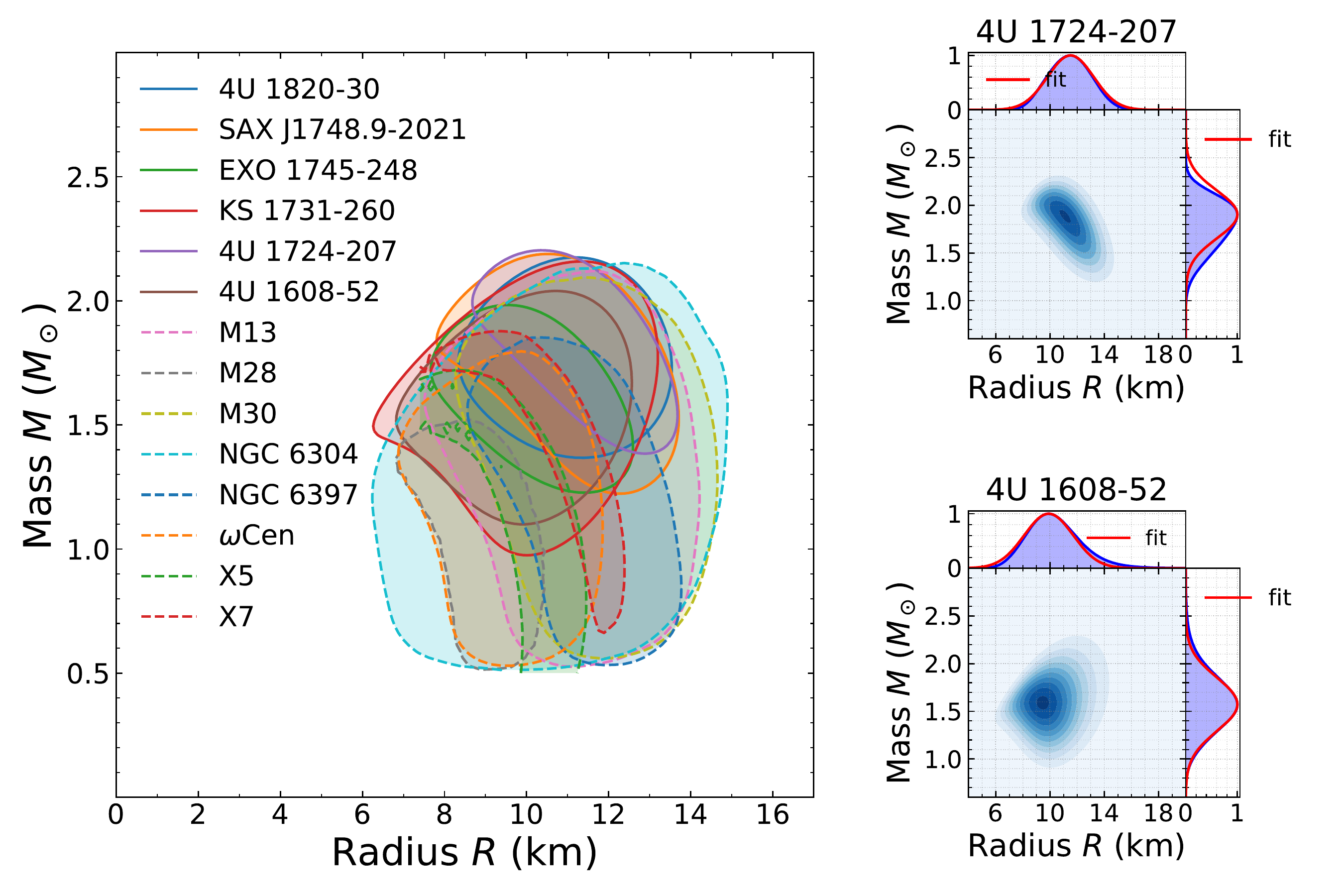}
\caption{(Left) Contour plot of the distributions of $M$ and $R$ for observed
  14 neutron stars.  The shaded regions are encircled by probability
  contours of $1\sigma$ (i.e., 68.27\%).\footnote{
  The original data is downloadable from
  \texttt{http://xtreme.as.arizona.edu/NeutronStars/}.}
  (Right) Two representatives of the neutron star data on the $R$-$M$ plane.}
\label{fig:contour_marginalize}
\end{figure}
%%%%%%%%%%%%%%%%%%%%%%%%%

Thanks to the recent advances in observations, the quality of the
neutron star observables is being improved steadily (see, e.g.,
Refs.~\cite{2016SPIE.9905E..1HG, TheLIGOScientific:2017qsa} for NICER
and GW170817).  To circumvent the above-mentioned difficulties,
current theoretical efforts are directed toward the EoS
inference from these stellar observables, especially masses $M$
and radii $R$ (pairs of which are called $M$-$R$ relation).
This is mediated by the Tolman--Oppenheimer--Volkoff (TOV) equation,
and the mapping from $M$-$R$ to the EoS is in one-to-one
correspondence~\cite{Lindblom:1992}.  Now, the Bayesian analysis
is one standard method to implement such inference~\cite{Ozel:2010fw,
  Steiner:2010fz, Steiner:2012xt, Raithel:2016bux, *Raithel:2017ity,
  Alvarez-Castillo:2016oln}.
If the number of available observational data is sufficiently large,
the likelihood would be well localized such that a choice
of the prior distribution scarcely affects the result.  In reality,
however, the number of data is limited, as tabulated in
Refs.~\cite{Ozel:2015fia, Ozel:2016oaf, Bogdanov:2016nle} and plotted in
Fig.~\ref{fig:contour_marginalize}~(Left), and we may not exclude such factors.
Hence, it would be complementary to develop an independent analysis
based on a different principle than the Bayesian analysis.

Here, we propose a new method to utilize the neural network in
the deep learning machinery to estimate the EoS from real
observational $M$-$R$ data, as an extension from
Ref.~\cite{Fujimoto:2017cdo}.  Deep learning provides us with a way to
find a regression function for complex nonlinear systems, and there
are many physics applications, which include QCD
physics~\cite{Pang:2016vdc,Mori:2017nwj}, nuclear
physics~\cite{Niu:2018csp}, and gravitational waves~\cite{George:2016hay}
(see also Ref.~\cite{Allen:2019dkq} and references therein).  As we explicate
below, an advantage to employ the deep learning method lies in the
fact that the numerical implementation is straightforward, so we are
relatively free from implicit biases.

%%%%%%%%%%%%%%%%%%%%%%%%%%%%%%%%%%%%%%%%%%%%%%%%%%
\section{Methods}
%%%%%%%%%%%%%%%%%%%%%%%%%%%%%%%%%%%%%%%%%%%%%%%%%%

%------------------------------------------------------------%
\subsection{Compilation of observational data}
%------------------------------------------------------------%

Ideally, with sufficient computational resources, machine learning would be
capable of directly dealing with full multidimensional data from the
observation.  Figure~\ref{fig:contour_marginalize}~(Left) shows only a
single contour for each neutron star, but the full data is available
in the form of the probability distribution as exemplified in
Fig.~\ref{fig:contour_marginalize}~(Right) for (arbitrarily chosen)
two representatives out of 14 observations.

In the present work we simplify our analysis by approximately
characterizing one probability distribution with four parameters.  We
project the two-dimensional distribution onto the one-dimensional
$M$-axis (and $R$-axis) by integrating over $R$ (and $M$,
respectively);  in other words, we make marginal distributions with
respect to $M$ and $R$.  Such marginal distributions are represented
by blue shaded shapes outside the frame on
Fig.~\ref{fig:contour_marginalize}~(Right).  Then, these two
distributions along the $M$-axis and the $R$-axis are fitted by
Gaussians as overlaid by red curves.  Since each Gaussian has two
parameters, namely, the mean and the variance, we sample $2\times
2\times 14=56$ parameters out from the raw $M$-$R$ data of 14 neutron
stars.  Now, our task is to find a mapping from these 56 observational
parameters onto the most likely EoS, i.e., $p=p(\rho)$.

%------------------------------------------------------------%
\subsection{Training and validation data with fluctuations}
%------------------------------------------------------------%

We will utilize the neural network to represent such a mapping, and for
the optimization, we generate training dataset;  many sets of randomly
generated EoS and the corresponding observational data.  It is
important to note that this mapping is not necessarily invertible;
even for the same EoS the observational data points may fluctuate
according to the probability distributions originating from
observational errors.  We need to train the neural network to
``recognize'' that the observational data points could depart from the
$M$-$R$ relation.

Here we outline how we generated the training and validation data for
our supervised learning.
The first step is the EoS
generation;  we divide a sufficiently wide density range,
$[\rho_0,\, 8\rho_0]$ in this work, into five segments equally
separated in the logarithmic scale, that is, $[\rho_{i-1},\, \rho_i]$
with $i=1,2,\dots 5$ and $\rho_5=8\rho_0$.  We then randomly choose an
average sound velocity in each segment, $c_{s,i}^2$, with a uniform
distribution in the causal range $\varepsilon < c_{s,i}^2 < 1
-\varepsilon$ (in the natural units $c=1$),
where we introduced a regulator, $\varepsilon=0.01$, to avoid singular behavior
in solving the TOV equations.  Note that the uniform distribution is
chosen to cover wide parameter regions efficiently.  Now we have 5
pressure values as $p_i=p_{i-1}+c_{s,i}^2(\rho_i-\rho_{i-1})$ for
$\rho=\rho_i$.

Up to $\rho = \rho_0$ we adopt a conventional nuclear EoS,
for which we chose SLy4~\cite{Douchin:2001sv}, one of the standard
EoSs for nuclear matter (meaning that $p_0=p(\rho_0)$ is fixed by
SLy4), and for $\rho > \rho_0$ the pressure is interpolated with a
polytrope function, i.e., $p=p(\rho)=K_i\, \rho^{\Gamma_i}$ for
$\rho_{i-1} < \rho < \rho_i$, where two parameters, $K_i$ and
$\Gamma_i$, are solved with two boundary conditions, $p_i=p(\rho_i)$
and $p_{i-1}=p(\rho_{i-1})$.

For a given EoS, the $M$-$R$ relation follows from the TOV equations,
which we call the \textit{genuine} $M$-$R$ curve.  We randomly sample
14 data points along the genuine $M$-$R$ curve in a region
$M>M_{\odot}$ (whose lower bound $M_{\odot}$ is chosen loosely so that
the region is large enough to cover masses from the actual observations).
Then, the variances of the Gaussian distribution, denoted by
$\sigma_{M_i}$ and $\sigma_{R_i}$, are randomly sampled from the
uniform distribution on the ranges $[0,\,M_{\odot})$ and
$[0,\,5\,\text{km})$, respectively.  These ranges are sufficient for
our purpose in view of Fig.~\ref{fig:contour_marginalize}.  The real
data are not necessarily centered on the \textit{bare} data point
($R_i^{(0)}$, $M_i^{(0)}$),
and we shall shift each distribution by $\Delta M_i$ and $\Delta R_i$
that we chose randomly from the Gaussian distributions with
$\sigma_{M_i}$ and $\sigma_{R_i}$.  To summarize the above, one
observation for the training data consists of 14 probability
distributions of the Gaussian shape whose center is
$(R_i^{(0)}+\Delta R_i,\, M_i^{(0)}+\Delta M_i)$
and variances are $\sigma_{R_i}$ and $\sigma_{M_i}$ along the $R$-axis
and the $M$-axis, respectively.
For the neural network to learn the
correlation between the variances ($\sigma_{R_i}$,\,$\sigma_{M_i}$)
and how far the actual data is off from the genuine $M$-$R$ curve,
we prepare 100 ensembles of different variances for each EoS and
sampled 14 data points, and then prepare 100 ensembles of shifts,
$\Delta M_i$ and $\Delta R_i$, for each generated set of variances.
This means that we prepared $100\times 100$ ensembles of data for each
EoS and sampled 14 data points.  For the training dataset we repeated
the above process 500 times to cover a wide variety of EoSs;  the
total training dataset is thus $100\times 100 \times 500 = 5,000,000$
sets of the EoS and the 14 data points.  For the validation dataset we
generate $1\times 1\times 100$ sets to monitor the convergence and
avoid the overfitting;  for each step of learning process, we
calculated the loss functions for the training and the validation data
(see Ref.~\cite{Fujimoto:2017cdo} for technicalities).

%------------------------------------------------------------%
\subsection{Neural network design}
%------------------------------------------------------------%

We specify the setups for the actual calculation.  For numerics 
we employ a Python library, Keras~\cite{software:Keras} using 
TensorFlow~\cite{arXiv:1605.08695} backend.
The design of our feedforward 
neural network is summarized in Tab.~\ref{tab:nnstructure}.
Our objective is to construct a network that can convert the
neutron star data to the EoS parameters, so the input and the output
layers have 56 and 5 neurons, respectively.  These correspond to 56
parameters of observed 14 neutron stars, ($M_i$, $R_i$,
$\sigma_{M,i}$, $\sigma_{R,i}$) ($i=1,2,\dots, 14$),
and 5 parameters of the EoS, $c_{s,i}^2$ ($i=1,2,\dots, 5$).
We chose the activation function at the output layer as 
$\sigma^{(4)}(x) = \tanh(x)$, so that the sound velocity automatically 
satisfies the causal bound.  For hidden layers the activation function 
is the ReLU, i.e., $\sigma^{(k)}(x) = \max\{0, x\}$ ($k=1,2,3$), which 
is known to evade the vanishing gradient problem and a standard
choice in deep learning~\cite{DBLP:journals/nature/LeCunBH15}.  We
implement the loss function by the mean square logarithmic errors
(\texttt{msle}).  The optimization method of our choice is
Adam~\cite{DBLP:journals/corr/KingmaB14} with the batch size 1000.  We
initialized neural network parameters with the Glorot uniform
distribution~\cite{pmlr-v9-glorot10a}.

%---   table   ---%
\begin{table}
  \centering 
  \begin{tabular}{lcc} \hline 
  \textbf{Layer} & \textbf{Number of neurons} & \textbf{Activation function} \\ \hline 
  0 (Input) & 56 & N/A \\ 
  1 & 60 & ReLU \\ 
  2, 3 & 40 & ReLU \\ 
  %% 3 & 40 & ReLU \\ 
  4 (Output) & 5  & $\tanh$ \\ \hline 
  \end{tabular}
  \caption{Our neural network architecture in this work.  In the input 
    layer 56 neurons correspond to parameters of 14 points of the 
    mass, the radius, and their variances.  In the output layer 5 
    neurons correspond to 5 parameters of the EoS.{}}
  \label{tab:nnstructure}
\end{table}
%---   table   ---%

%------------------------------------------------------------%
\subsection{Uncertainty estimate from credibility of reproducibility}
\label{sec:cor}
%------------------------------------------------------------%

In our strategy we took care of the probability distribution in the
observational side only, but the deduced EoS also has such a
probability distribution around the most likely curve.  To implement
that, instead of randomly generating EoSs, we could have generated
some distributions on the $\rho$-$p$ plane and sample fluctuating EoSs
according to the generated distribution, which would, however,
increase the size of the training dataset tens of thousands larger and
require gigantic computational resources.

Here, we employ an alternative practical way to quantify the
credibility of the deduced EoS with less efforts.  We generate 10
independent training datasets to prepare 10 independent neural
network models.  For the same real experimental data, those 10 neural
network models output 10 deduced EoSs.  If a part of the EoS is
insensitive to the $M$-$R$ observation, different neural network
models would lead to different EoSs in such an unconstrained region.
From the dispersion over 10 deduced EoSs, therefore, we can estimate
the credibility of our results.  Strictly speaking, this dispersion is
not the probability distribution of the likely EoS but a measure to
quantify how much the same deduced EoS is reproduced with the same
analysis.  In other words, this measure is to be regarded as the
credibility of reproducibility within the present setup of machine
learning.  If the physical error bar is large, the credibility band
would be large, but a small credibility band does not always guarantee
small physical error bar.  In this sense our uncertainty estimate
gives a lower bound.
Here we note that the uncertainty estimated in this way accounts for
the statistical part (see the band labeled by ``10 NNs'' in
Figs.~\ref{fig:resulteos} and \ref{fig:resultmr}).  Uncertainty
including systematics can be quantified by the root-mean-square
deviation between the guessed and true values using the validation
data (see the band labeled by ``validation'' in
Figs.~\ref{fig:resulteos} and \ref{fig:resultmr}), as addressed in
Ref.~\cite{Fujimoto:2017cdo}.  This leads to an uncertainty width of
1.7\,km for $R$ at $M=1.4M_\odot$ in the $M$-$R$ plane, which is
comparable to our inferred width of 1.3\,km (68\% CL).

%%%%%%%%%%%%%%%%%%%%%%%%%%%%%%%%%%%%%%%%%%%%%%%%%%
\section{Results and discussions}
%%%%%%%%%%%%%%%%%%%%%%%%%%%%%%%%%%%%%%%%%%%%%%%%%%

%------------------------------------------------------------%
\subsection{Deduced equation of state}
%------------------------------------------------------------%

%%%%%%%%%% FIGURE %%%%%%%%%%
\begin{figure}
\centering
\includegraphics[width=0.7\textwidth]{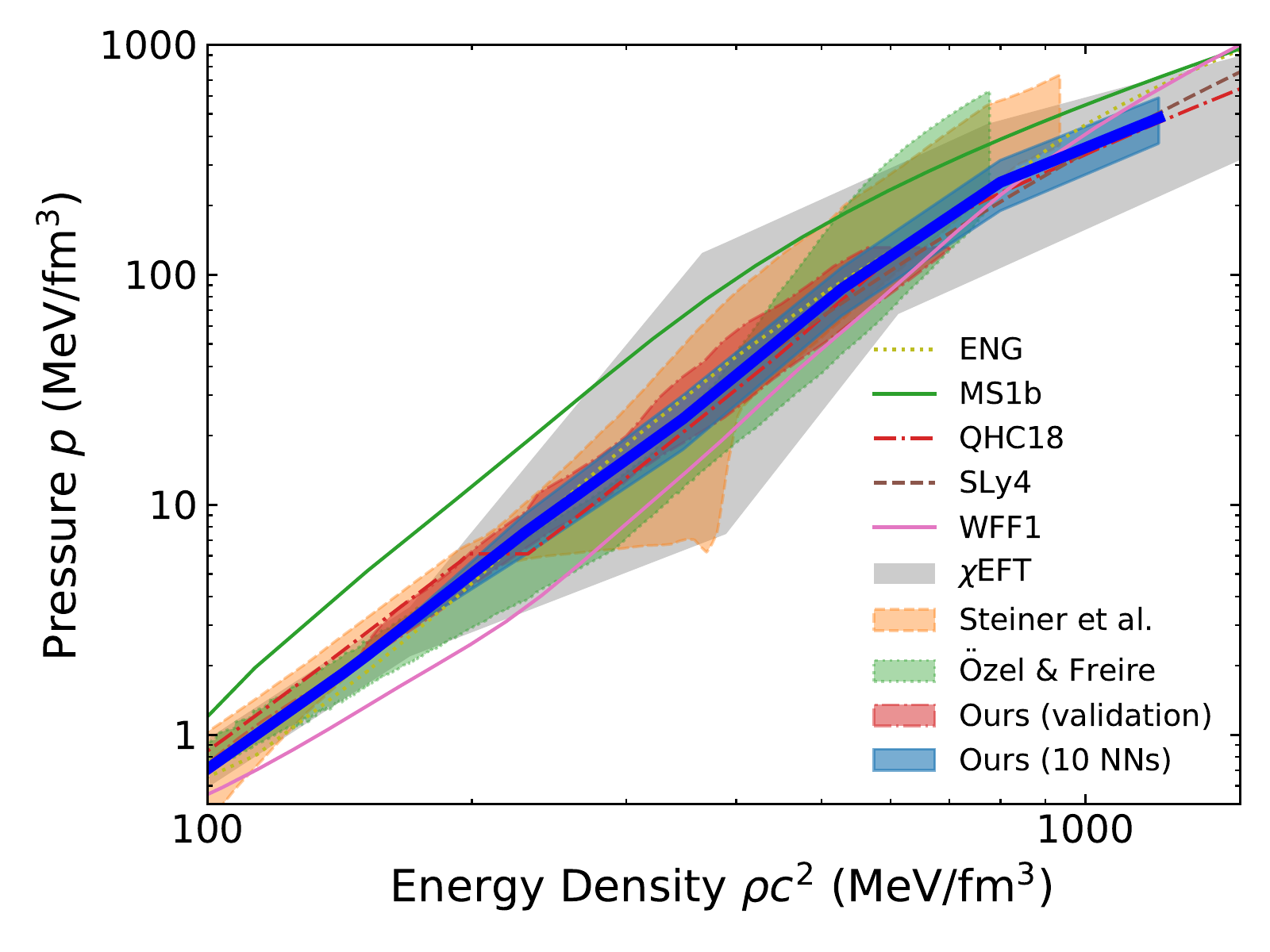}
\caption{EoS (``Ours'' drawn by blue line) deduced from the
  experimental data of 14 neutron stars as shown in
  Fig.~\ref{fig:contour_marginalize}.  The light red and blue shades
  represent our 68\% credibility band (``validation'' and ``10 NNs'')
  evaluated in different ways; see Sec.~\ref{sec:cor} for the precise
  meaning.  Phenomenological EoS candidates, the $\chi$EFT prediction
  and results inferred from Bayesian methods (Steiner \textit{et
    al.}~\cite{Steiner:2012xt} and {\"Ozel} \&
  Freire~\cite{Ozel:2016oaf}) are overlaid for reference.  The
  former~\cite{Steiner:2012xt} represents 68\% CL, and the
  latter~\cite{Ozel:2016oaf} shows the contour of $e^{-1}$ of the
  maximum likelihood.}
\label{fig:resulteos}
\end{figure}
%%%%%%%%%%%%%%%%%%%%%%%%%

In Fig.~\ref{fig:resulteos} we present the deduced EoS by the blue
line and its credibility by the light blue shade (labeled by ``10
NNs'').
Uncertainty quantified in a different way is also overlaid by the
light red shade in Fig.~\ref{fig:resulteos} (labeled by
``validation'').  Our results are in favor of standard EoSs calculated
within the nuclear many-body model, such as APR4~\cite{Akmal:1998cf},
BSk20~\cite{Goriely:2010bm}, ENG (Dirac--Brueckner--Hartree--Fock
method)~\cite{Engvik:1995gn}, and SLy4 (non-relativistic
potential)~\cite{Douchin:2001sv}, some of which are overlaid on
Fig.~\ref{fig:resulteos}.  Our results indicate that the constraints
from currently observed neutron stars do not have enough resolution to
probe a possibility of the first-order phase transition as encoded in
QHC18 (hybrid phenomenological construction)~\cite{Baym:2017whm}.  The
gray band represents an estimate from the chiral effective theory
($\chi$EFT)~\cite{Hebeler:2013nza}, and our results lie within this
band.  In Fig.~\ref{fig:resulteos}, for reference, we show MS1b (relativistic mean-field)~\cite{Mueller:1996pm}, WFF1 (variational)~\cite{Wiringa:1988tp}, and several other
phenomenological EoSs.
The Bayesian analyses~\cite{Steiner:2012xt, Ozel:2016oaf} are also
overlaid in Fig.~\ref{fig:resulteos}.  Note that while {\"Ozel} \&
Freire~\cite{Ozel:2016oaf} and our present analysis use the same
astrophysical data, Steiner \textit{et al.}~\cite{Steiner:2012xt}
employs eight X-ray sources.

\begin{figure}
\centering
\includegraphics[width=0.7\textwidth]{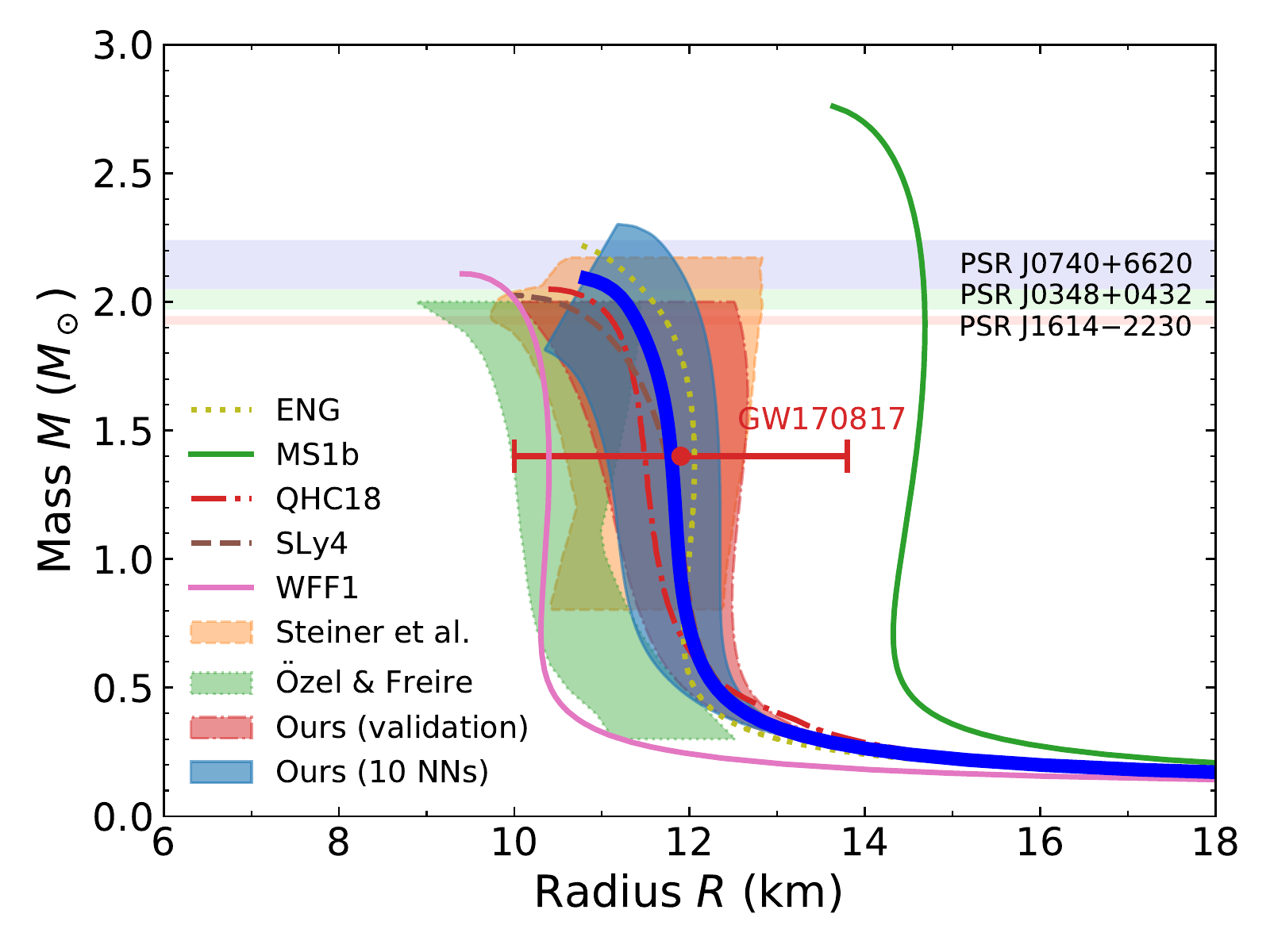}
\caption{$M$-$R$ relations corresponding to the deduced EoS (Ours) with
  phenomenological EoS candidates and Bayesian analyses (Steiner \textit{et
  al.}\@ and {\"Ozel} \& Freire) as shown in Fig.~\ref{fig:resulteos}.}
\label{fig:resultmr}
\end{figure}

It is an interesting question how the corresponding $M$-$R$ curve
looks like because even the knowledge of the existence of the $M$-$R$ curve
is not provided to the neural network during the supervised learning
by the $M$-$R$ points and the EoS parameters.
Figure~\ref{fig:resultmr} shows the $M$-$R$ curves
corresponding to the EoSs in Fig.~\ref{fig:resulteos}.  We see that
our deduced EoS (blue curve) certainly supports massive neutron stars
above two solar mass~\cite{Demorest:2010bx, *Fonseca:2016tux,
  Antoniadis:2013pzd, Cromartie:2019kug}.

%%%%%%%%%% FIGURE %%%%%%%%%%
%%%%%%%%%%%%%%%%%%%%%%%%%

%------------------------------------------------------------%
\subsection{Discussions}
%------------------------------------------------------------%

One may want to know why the uncertainty band of our deduced EoS looks
such narrow.  A part of the reason lies in the boundary condition in
the low density side;  we assumed SLy4 for $\rho \leq \rho_0$ because
up to this density the EoS is well constrained by nuclear properties
accessible by terrestrial experiments.
So our results should be more precisely regarded as the most likely
extrapolation from SLy4 with help of the observational data of 14
neutron stars.
It shall be a future work to inspect possible bias
effect induced by such a choice of the EoS up to $\rho_0$.
Also we can in principle remove such an assumption by extending
the neural network architecture including data from nuclear physics
experiments (e.g., symmetry energy; see discussions in
Ref.~\cite{Gandolfi:2011xu}) on top of neutron star data.  Such a
global analysis over all available data from astrophysics and nuclear
physics experiments would be an ambitious future challenge.

At the same time, we can argue from a different point of view.  The
light blue band in Fig.~\ref{fig:resulteos} may look small at first
glance, but the resolution is not yet good enough to justify/falsify a
first-order phase transition.  In view of the light blue band
in Fig.~\ref{fig:resultmr}, the corresponding uncertainty for the
$M$-$R$ relation is $\sim 1\,\text{km}$.
%% does not appear so small.

%%%%%%%%%% FIGURE %%%%%%%%%%
\begin{figure}
 \begin{minipage}[t]{0.47\textwidth}
\centering
\includegraphics[width=\textwidth]{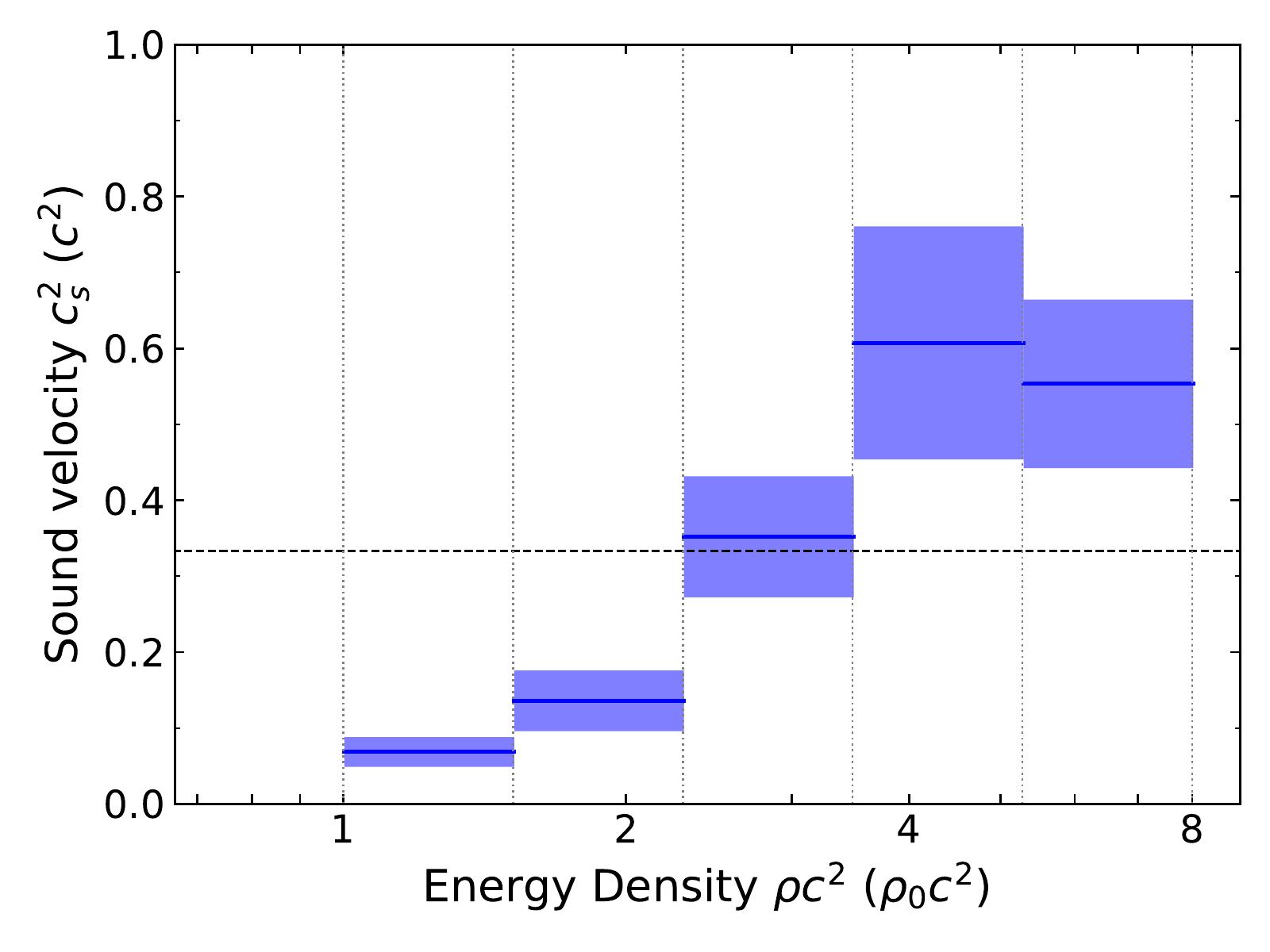}
\caption{Sound velocity in each segment corresponding to the deduced
  EoS in Fig.~\ref{fig:resulteos}.  The band represents 68\%
  credibility.  The horizontal dotted line
  represents the conformal limit of $c_s = 1/\sqrt{3}$.}
\label{fig:sound}
 \end{minipage}
 \hspace{1em}
 \begin{minipage}[t]{0.47\textwidth}
\centering 
\includegraphics[width=\textwidth]{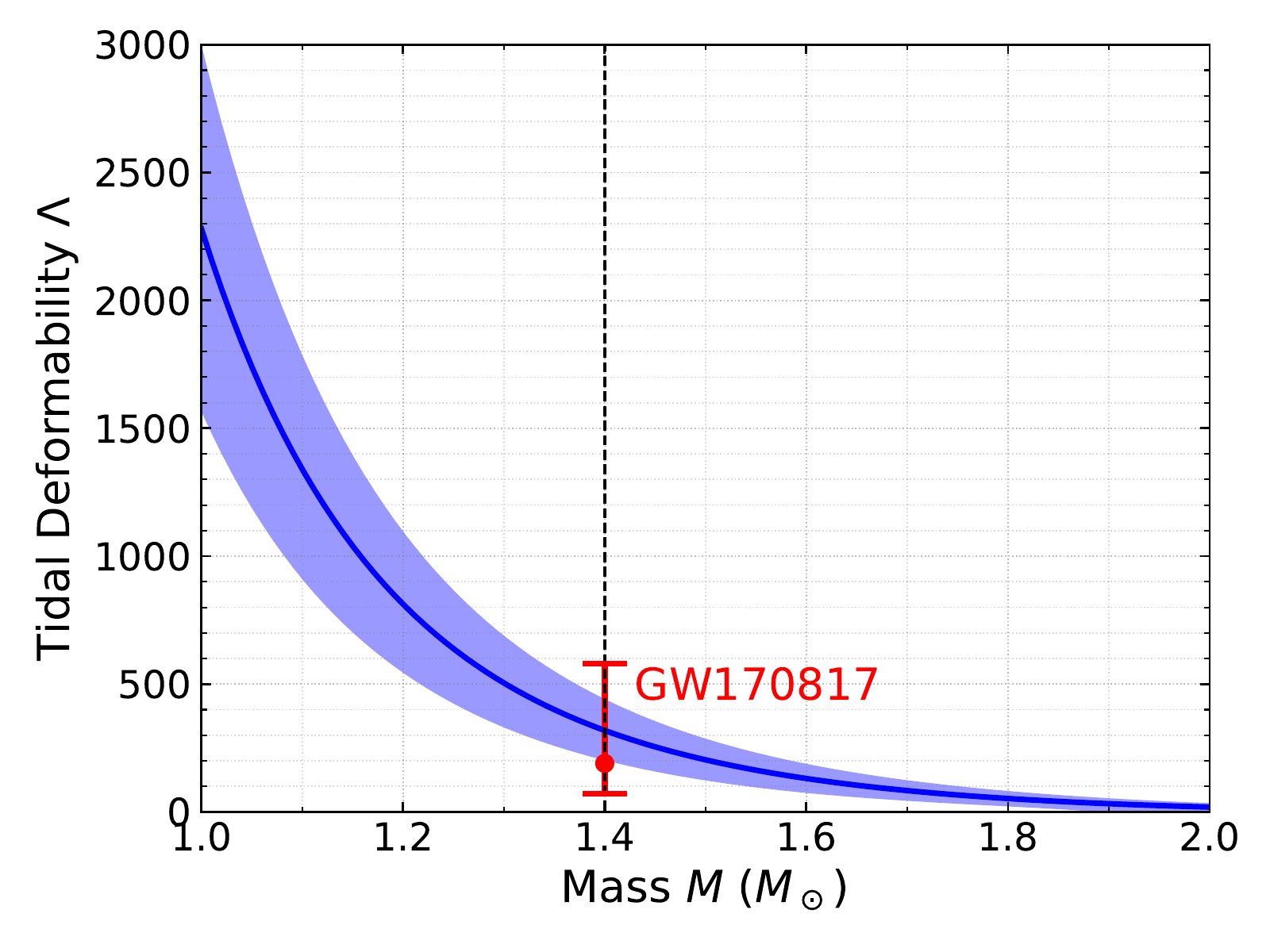}
\caption{Tidal deformability $\Lambda$ from the deduced EoS and the
  experimental bound from GW170817 (red bar).  The band represents
  68\% credibility.}
\label{fig:tidal}
 \end{minipage}
\end{figure}
%%%%%%%%%%%%%%%%%%%%%%%%%

Another important physical quantity derived from the EoS is the
sound velocity, $c_s$, which is plotted in Fig.~\ref{fig:sound}.
Interestingly, the deduced sound velocity is smaller than $1/\sqrt{3}$
(the conformal limit value, viz., a naive upper bound for massless
ultrarelativistic systems) for $\rho \lesssim 2\rho_0$.
With further increasing $\rho > 2\rho_0$, the sound velocity becomes
significantly greater than $1/\sqrt{3}$, and eventually the increasing
behavior is saturated beyond $\sim 4\rho_0$.  Such a sharp increase of
the sound velocity around $2\rho_0$ appears in accordance with the
recent studies~\cite{Bedaque:2014sqa, Tews:2018kmu}.  At even higher
densities $>4\rho_0$ it is likely that the sound velocity starts
decreasing and approaches the conformal limit of asymptotically free
quarks and gluons.  This in turn implies that the saturation seen
around $4\rho_0$ hints a transition to weakly interacting quark
matter.

Finally, we shall confirm that our deduced EoS is consistent with the
recent gravitational wave experiment, specifically the tidal
deformability $\Lambda$.  Once the EoS is given, the tidal
deformability can be calculated following the method outlined in the
Ref.~\cite{Hinderer:2009ca}.  The experimentally
determined bound,
$\Lambda(1.4M_{\odot})=190^{+390}_{-120}$~\cite{Abbott:2018exr},
is indicated by a red bar in Fig.~\ref{fig:tidal}.  Our deduced EoS
leads to $\Lambda(1.4\Msun) = 320 \pm 120$ which is entirely
consistent with the GW170817 measurement within the error bar as it
should be.
For the moment we utilize the tidal deformability as a benchmark test,
but in the future the neural network should be better designed to
implement what is called the multi-messenger observation, inclusive of
gravitational waves as well as electromagnetic waves.

%%%%%%%%%%%%%%%%%%%%%%%%%%%%%%%%%%%%%%%%%%%%%%%%%%
\section{Summary}
%%%%%%%%%%%%%%%%%%%%%%%%%%%%%%%%%%%%%%%%%%%%%%%%%%

In this work we successfully utilized a new method based on the
machine learning to infer neutron star EoS in a way independent of the
existing methods.
In our method the deep neural network can deal with
nonlinear mapping from masses $M$ and radii $R$ of neutron stars to
the EoS parameters.  The neural network model was optimized with training
datasets of size 5,000,000, and the convergence was monitored with an
independent validation dataset.  In this way, from available $M$-$R$
data from 14 neutron stars, we deduced an EoS to find it compatible
with the conventional nuclear EoS and the currently existing
constraints.
Dealing with two-dimensional $M$-$R$ distribution for the neural
network input would be an important extension for the future. Still,
our successful results would be a first step toward further
refinements to incorporate the gravitational wave measurements and
nuclear physics experiments.  Machine learning's advantage lies in
handling such a large set of complex data, and this direction deserves
investigations.

\begin{acknowledgments}

We thank Toru~Kojo, Andrew~Steiner and Wolfram~Weise for encouraging
discussions.  K.~F.\ was supported by Japan Society for the Promotion
of Science (JSPS) KAKENHI Grant No. 18H01211.

\end{acknowledgments}

\bibliographystyle{apsrev4-1}
\bibliography{ref_deep}

\begin{thebibliography}{52}%
\makeatletter
\providecommand \@ifxundefined [1]{%
 \@ifx{#1\undefined}
}%
\providecommand \@ifnum [1]{%
 \ifnum #1\expandafter \@firstoftwo
 \else \expandafter \@secondoftwo
 \fi
}%
\providecommand \@ifx [1]{%
 \ifx #1\expandafter \@firstoftwo
 \else \expandafter \@secondoftwo
 \fi
}%
\providecommand \natexlab [1]{#1}%
\providecommand \enquote  [1]{``#1''}%
\providecommand \bibnamefont  [1]{#1}%
\providecommand \bibfnamefont [1]{#1}%
\providecommand \citenamefont [1]{#1}%
\providecommand \href@noop [0]{\@secondoftwo}%
\providecommand \href [0]{\begingroup \@sanitize@url \@href}%
\providecommand \@href[1]{\@@startlink{#1}\@@href}%
\providecommand \@@href[1]{\endgroup#1\@@endlink}%
\providecommand \@sanitize@url [0]{\catcode `\\12\catcode `\$12\catcode
  `\&12\catcode `\#12\catcode `\^12\catcode `\_12\catcode `\%12\relax}%
\providecommand \@@startlink[1]{}%
\providecommand \@@endlink[0]{}%
\providecommand \url  [0]{\begingroup\@sanitize@url \@url }%
\providecommand \@url [1]{\endgroup\@href {#1}{\urlprefix }}%
\providecommand \urlprefix  [0]{URL }%
\providecommand \Eprint [0]{\href }%
\providecommand \doibase [0]{http://dx.doi.org/}%
\providecommand \selectlanguage [0]{\@gobble}%
\providecommand \bibinfo  [0]{\@secondoftwo}%
\providecommand \bibfield  [0]{\@secondoftwo}%
\providecommand \translation [1]{[#1]}%
\providecommand \BibitemOpen [0]{}%
\providecommand \bibitemStop [0]{}%
\providecommand \bibitemNoStop [0]{.\EOS\space}%
\providecommand \EOS [0]{\spacefactor3000\relax}%
\providecommand \BibitemShut  [1]{\csname bibitem#1\endcsname}%
\let\auto@bib@innerbib\@empty
%</preamble>
\bibitem [{\citenamefont {Haensel}\ \emph {et~al.}(2007)\citenamefont
  {Haensel}, \citenamefont {Potekhin},\ and\ \citenamefont
  {Yakovlev}}]{Haensel:2007yy}%
  \BibitemOpen
  \bibfield  {author} {\bibinfo {author} {\bibfnamefont {P.}~\bibnamefont
  {Haensel}}, \bibinfo {author} {\bibfnamefont {A.~Y.}\ \bibnamefont
  {Potekhin}}, \ and\ \bibinfo {author} {\bibfnamefont {D.~G.}\ \bibnamefont
  {Yakovlev}},\ }\href {\doibase 10.1007/978-0-387-47301-7} {\bibfield
  {journal} {\bibinfo  {journal} {Astrophys. Space Sci. Libr.}\ }\textbf
  {\bibinfo {volume} {326}},\ \bibinfo {pages} {pp.1} (\bibinfo {year}
  {2007})}\BibitemShut {NoStop}%
%%CITATION = ASSLA,326,pp.1;%%
\bibitem [{\citenamefont {Lattimer}(2012)}]{Lattimer:2012nd}%
  \BibitemOpen
  \bibfield  {author} {\bibinfo {author} {\bibfnamefont {J.~M.}\ \bibnamefont
  {Lattimer}},\ }\href {\doibase 10.1146/annurev-nucl-102711-095018} {\bibfield
   {journal} {\bibinfo  {journal} {Ann. Rev. Nucl. Part. Sci.}\ }\textbf
  {\bibinfo {volume} {62}},\ \bibinfo {pages} {485} (\bibinfo {year} {2012})},\
  \Eprint {http://arxiv.org/abs/1305.3510} {arXiv:1305.3510 [nucl-th]}
  \BibitemShut {NoStop}%
%%CITATION = ARXIV:1305.3510;%%
\bibitem [{\citenamefont {{\"Ozel}}\ and\ \citenamefont
  {Freire}(2016)}]{Ozel:2016oaf}%
  \BibitemOpen
  \bibfield  {author} {\bibinfo {author} {\bibfnamefont {F.}~\bibnamefont
  {{\"Ozel}}}\ and\ \bibinfo {author} {\bibfnamefont {P.}~\bibnamefont
  {Freire}},\ }\href {\doibase 10.1146/annurev-astro-081915-023322} {\bibfield
  {journal} {\bibinfo  {journal} {Ann. Rev. Astron. Astrophys.}\ }\textbf
  {\bibinfo {volume} {54}},\ \bibinfo {pages} {401} (\bibinfo {year} {2016})},\
  \Eprint {http://arxiv.org/abs/1603.02698} {arXiv:1603.02698 [astro-ph.HE]}
  \BibitemShut {NoStop}%
%%CITATION = ARXIV:1603.02698;%%
\bibitem [{\citenamefont {Baym}\ \emph {et~al.}(2018)\citenamefont {Baym},
  \citenamefont {Hatsuda}, \citenamefont {Kojo}, \citenamefont {Powell},
  \citenamefont {Song},\ and\ \citenamefont {Takatsuka}}]{Baym:2017whm}%
  \BibitemOpen
  \bibfield  {author} {\bibinfo {author} {\bibfnamefont {G.}~\bibnamefont
  {Baym}}, \bibinfo {author} {\bibfnamefont {T.}~\bibnamefont {Hatsuda}},
  \bibinfo {author} {\bibfnamefont {T.}~\bibnamefont {Kojo}}, \bibinfo {author}
  {\bibfnamefont {P.~D.}\ \bibnamefont {Powell}}, \bibinfo {author}
  {\bibfnamefont {Y.}~\bibnamefont {Song}}, \ and\ \bibinfo {author}
  {\bibfnamefont {T.}~\bibnamefont {Takatsuka}},\ }\href {\doibase
  10.1088/1361-6633/aaae14} {\bibfield  {journal} {\bibinfo  {journal} {Rept.
  Prog. Phys.}\ }\textbf {\bibinfo {volume} {81}},\ \bibinfo {pages} {056902}
  (\bibinfo {year} {2018})},\ \Eprint {http://arxiv.org/abs/1707.04966}
  {arXiv:1707.04966 [astro-ph.HE]} \BibitemShut {NoStop}%
%%CITATION = ARXIV:1707.04966;%%
\bibitem [{\citenamefont {Blaschke}\ and\ \citenamefont
  {Chamel}(2018)}]{Blaschke:2018mqw}%
  \BibitemOpen
  \bibfield  {author} {\bibinfo {author} {\bibfnamefont {D.}~\bibnamefont
  {Blaschke}}\ and\ \bibinfo {author} {\bibfnamefont {N.}~\bibnamefont
  {Chamel}},\ }\href@noop {} {\  (\bibinfo {year} {2018})},\ \Eprint
  {http://arxiv.org/abs/1803.01836} {arXiv:1803.01836 [nucl-th]} \BibitemShut
  {NoStop}%
%%CITATION = ARXIV:1803.01836;%%
\bibitem [{\citenamefont {{Sch\"afer}}\ and\ \citenamefont
  {Wilczek}(1999)}]{Schafer:1998ef}%
  \BibitemOpen
  \bibfield  {author} {\bibinfo {author} {\bibfnamefont {T.}~\bibnamefont
  {{Sch\"afer}}}\ and\ \bibinfo {author} {\bibfnamefont {F.}~\bibnamefont
  {Wilczek}},\ }\href {\doibase 10.1103/PhysRevLett.82.3956} {\bibfield
  {journal} {\bibinfo  {journal} {Phys. Rev. Lett.}\ }\textbf {\bibinfo
  {volume} {82}},\ \bibinfo {pages} {3956} (\bibinfo {year} {1999})},\ \Eprint
  {http://arxiv.org/abs/hep-ph/9811473} {arXiv:hep-ph/9811473 [hep-ph]}
  \BibitemShut {NoStop}%
%%CITATION = HEP-PH/9811473;%%
\bibitem [{\citenamefont {McLerran}\ and\ \citenamefont
  {Pisarski}(2007)}]{McLerran:2007qj}%
  \BibitemOpen
  \bibfield  {author} {\bibinfo {author} {\bibfnamefont {L.}~\bibnamefont
  {McLerran}}\ and\ \bibinfo {author} {\bibfnamefont {R.~D.}\ \bibnamefont
  {Pisarski}},\ }\href {\doibase 10.1016/j.nuclphysa.2007.08.013} {\bibfield
  {journal} {\bibinfo  {journal} {Nucl. Phys.}\ }\textbf {\bibinfo {volume}
  {A796}},\ \bibinfo {pages} {83} (\bibinfo {year} {2007})},\ \Eprint
  {http://arxiv.org/abs/0706.2191} {arXiv:0706.2191 [hep-ph]} \BibitemShut
  {NoStop}%
%%CITATION = ARXIV:0706.2191;%%
\bibitem [{\citenamefont {Fukushima}\ and\ \citenamefont
  {Kojo}(2016)}]{Fukushima:2015bda}%
  \BibitemOpen
  \bibfield  {author} {\bibinfo {author} {\bibfnamefont {K.}~\bibnamefont
  {Fukushima}}\ and\ \bibinfo {author} {\bibfnamefont {T.}~\bibnamefont
  {Kojo}},\ }\href {\doibase 10.3847/0004-637X/817/2/180} {\bibfield  {journal}
  {\bibinfo  {journal} {Astrophys. J.}\ }\textbf {\bibinfo {volume} {817}},\
  \bibinfo {pages} {180} (\bibinfo {year} {2016})},\ \Eprint
  {http://arxiv.org/abs/1509.00356} {arXiv:1509.00356 [nucl-th]} \BibitemShut
  {NoStop}%
%%CITATION = ARXIV:1509.00356;%%
\bibitem [{\citenamefont {McLerran}\ and\ \citenamefont
  {Reddy}(2018)}]{McLerran:2018hbz}%
  \BibitemOpen
  \bibfield  {author} {\bibinfo {author} {\bibfnamefont {L.}~\bibnamefont
  {McLerran}}\ and\ \bibinfo {author} {\bibfnamefont {S.}~\bibnamefont
  {Reddy}},\ }\href@noop {} {\  (\bibinfo {year} {2018})},\ \Eprint
  {http://arxiv.org/abs/1811.12503} {arXiv:1811.12503 [nucl-th]} \BibitemShut
  {NoStop}%
%%CITATION = ARXIV:1811.12503;%%
\bibitem [{\citenamefont {Masuda}\ \emph {et~al.}(2013)\citenamefont {Masuda},
  \citenamefont {Hatsuda},\ and\ \citenamefont {Takatsuka}}]{Masuda:2012kf}%
  \BibitemOpen
  \bibfield  {author} {\bibinfo {author} {\bibfnamefont {K.}~\bibnamefont
  {Masuda}}, \bibinfo {author} {\bibfnamefont {T.}~\bibnamefont {Hatsuda}}, \
  and\ \bibinfo {author} {\bibfnamefont {T.}~\bibnamefont {Takatsuka}},\ }\href
  {\doibase 10.1088/0004-637X/764/1/12} {\bibfield  {journal} {\bibinfo
  {journal} {Astrophys. J.}\ }\textbf {\bibinfo {volume} {764}},\ \bibinfo
  {pages} {12} (\bibinfo {year} {2013})},\ \Eprint
  {http://arxiv.org/abs/1205.3621} {arXiv:1205.3621 [nucl-th]} \BibitemShut
  {NoStop}%
%%CITATION = ARXIV:1205.3621;%%
\bibitem [{\citenamefont {Alvarez-Castillo}\ \emph {et~al.}(2014)\citenamefont
  {Alvarez-Castillo}, \citenamefont {Benic}, \citenamefont {Blaschke},\ and\
  \citenamefont {\L{}astowiecki}}]{Alvarez-Castillo:2013spa}%
  \BibitemOpen
  \bibfield  {author} {\bibinfo {author} {\bibfnamefont {D.~E.}\ \bibnamefont
  {Alvarez-Castillo}}, \bibinfo {author} {\bibfnamefont {S.}~\bibnamefont
  {Benic}}, \bibinfo {author} {\bibfnamefont {D.}~\bibnamefont {Blaschke}}, \
  and\ \bibinfo {author} {\bibfnamefont {R.}~\bibnamefont {\L{}astowiecki}},\
  }\href {\doibase 10.5506/APhysPolBSupp.7.203} {\bibfield  {journal} {\bibinfo
   {journal} {Acta Phys. Polon. Supp.}\ }\textbf {\bibinfo {volume} {7}},\
  \bibinfo {pages} {203} (\bibinfo {year} {2014})},\ \Eprint
  {http://arxiv.org/abs/1311.5112} {arXiv:1311.5112 [nucl-th]} \BibitemShut
  {NoStop}%
%%CITATION = ARXIV:1311.5112;%%
\bibitem [{\citenamefont {Oter}\ \emph {et~al.}(2019)\citenamefont {Oter},
  \citenamefont {Windisch}, \citenamefont {Llanes-Estrada},\ and\ \citenamefont
  {Alford}}]{Oter:2019kig}%
  \BibitemOpen
  \bibfield  {author} {\bibinfo {author} {\bibfnamefont {E.~L.}\ \bibnamefont
  {Oter}}, \bibinfo {author} {\bibfnamefont {A.}~\bibnamefont {Windisch}},
  \bibinfo {author} {\bibfnamefont {F.~J.}\ \bibnamefont {Llanes-Estrada}}, \
  and\ \bibinfo {author} {\bibfnamefont {M.}~\bibnamefont {Alford}},\
  }\href@noop {} {\  (\bibinfo {year} {2019})},\ \Eprint
  {http://arxiv.org/abs/1901.05271} {arXiv:1901.05271 [gr-qc]} \BibitemShut
  {NoStop}%
%%CITATION = ARXIV:1901.05271;%%
\bibitem [{\citenamefont {Aarts}(2016)}]{Aarts:2015tyj}%
  \BibitemOpen
  \bibfield  {author} {\bibinfo {author} {\bibfnamefont {G.}~\bibnamefont
  {Aarts}},\ }\href {\doibase 10.1088/1742-6596/706/2/022004} {\bibfield
  {journal} {\bibinfo  {journal} {J. Phys. Conf. Ser.}\ }\textbf {\bibinfo
  {volume} {706}},\ \bibinfo {pages} {022004} (\bibinfo {year} {2016})},\
  \Eprint {http://arxiv.org/abs/1512.05145} {arXiv:1512.05145 [hep-lat]}
  \BibitemShut {NoStop}%
%%CITATION = ARXIV:1512.05145;%%
\bibitem [{\citenamefont {Kurkela}\ \emph {et~al.}(2010)\citenamefont
  {Kurkela}, \citenamefont {Romatschke},\ and\ \citenamefont
  {Vuorinen}}]{Kurkela:2009gj}%
  \BibitemOpen
  \bibfield  {author} {\bibinfo {author} {\bibfnamefont {A.}~\bibnamefont
  {Kurkela}}, \bibinfo {author} {\bibfnamefont {P.}~\bibnamefont {Romatschke}},
  \ and\ \bibinfo {author} {\bibfnamefont {A.}~\bibnamefont {Vuorinen}},\
  }\href {\doibase 10.1103/PhysRevD.81.105021} {\bibfield  {journal} {\bibinfo
  {journal} {Phys. Rev.}\ }\textbf {\bibinfo {volume} {D81}},\ \bibinfo {pages}
  {105021} (\bibinfo {year} {2010})},\ \Eprint {http://arxiv.org/abs/0912.1856}
  {arXiv:0912.1856 [hep-ph]} \BibitemShut {NoStop}%
%%CITATION = ARXIV:0912.1856;%%
\bibitem [{\citenamefont {Gendreau}\ \emph {et~al.}(2016)\citenamefont
  {Gendreau} \emph {et~al.}}]{2016SPIE.9905E..1HG}%
  \BibitemOpen
  \bibfield  {author} {\bibinfo {author} {\bibfnamefont {K.~C.}\ \bibnamefont
  {Gendreau}} \emph {et~al.},\ }in\ \href {\doibase 10.1117/12.2231304} {\emph
  {\bibinfo {booktitle} {Space Telescopes and Instrumentation 2016: Ultraviolet
  to Gamma Ray}}},\ \bibinfo {series} {Proc. SPIE}, Vol.\ \bibinfo {volume}
  {9905}\ (\bibinfo {year} {2016})\ p.\ \bibinfo {pages} {99051H}\BibitemShut
  {NoStop}%
\bibitem [{\citenamefont {Abbott}\ \emph {et~al.}(2017)\citenamefont {Abbott}
  \emph {et~al.}}]{TheLIGOScientific:2017qsa}%
  \BibitemOpen
  \bibfield  {author} {\bibinfo {author} {\bibfnamefont {B.~P.}\ \bibnamefont
  {Abbott}} \emph {et~al.} (\bibinfo {collaboration} {Virgo, LIGO
  Scientific}),\ }\href {\doibase 10.1103/PhysRevLett.119.161101} {\bibfield
  {journal} {\bibinfo  {journal} {Phys. Rev. Lett.}\ }\textbf {\bibinfo
  {volume} {119}},\ \bibinfo {pages} {161101} (\bibinfo {year} {2017})},\
  \Eprint {http://arxiv.org/abs/1710.05832} {arXiv:1710.05832 [gr-qc]}
  \BibitemShut {NoStop}%
%%CITATION = ARXIV:1710.05832;%%
\bibitem [{\citenamefont {Lindblom}(1992)}]{Lindblom:1992}%
  \BibitemOpen
  \bibfield  {author} {\bibinfo {author} {\bibfnamefont {L.}~\bibnamefont
  {Lindblom}},\ }\href {\doibase 10.1086/171882} {\bibfield  {journal}
  {\bibinfo  {journal} {Astrophys. J.}\ }\textbf {\bibinfo {volume} {398}},\
  \bibinfo {pages} {569} (\bibinfo {year} {1992})}\BibitemShut {NoStop}%
\bibitem [{\citenamefont {{\"Ozel}}\ \emph {et~al.}(2010)\citenamefont
  {{\"Ozel}}, \citenamefont {Baym},\ and\ \citenamefont {Guver}}]{Ozel:2010fw}%
  \BibitemOpen
  \bibfield  {author} {\bibinfo {author} {\bibfnamefont {F.}~\bibnamefont
  {{\"Ozel}}}, \bibinfo {author} {\bibfnamefont {G.}~\bibnamefont {Baym}}, \
  and\ \bibinfo {author} {\bibfnamefont {T.}~\bibnamefont {Guver}},\ }\href
  {\doibase 10.1103/PhysRevD.82.101301} {\bibfield  {journal} {\bibinfo
  {journal} {Phys. Rev.}\ }\textbf {\bibinfo {volume} {D82}},\ \bibinfo {pages}
  {101301} (\bibinfo {year} {2010})},\ \Eprint {http://arxiv.org/abs/1002.3153}
  {arXiv:1002.3153 [astro-ph.HE]} \BibitemShut {NoStop}%
%%CITATION = ARXIV:1002.3153;%%
\bibitem [{\citenamefont {Steiner}\ \emph {et~al.}(2010)\citenamefont
  {Steiner}, \citenamefont {Lattimer},\ and\ \citenamefont
  {Brown}}]{Steiner:2010fz}%
  \BibitemOpen
  \bibfield  {author} {\bibinfo {author} {\bibfnamefont {A.~W.}\ \bibnamefont
  {Steiner}}, \bibinfo {author} {\bibfnamefont {J.~M.}\ \bibnamefont
  {Lattimer}}, \ and\ \bibinfo {author} {\bibfnamefont {E.~F.}\ \bibnamefont
  {Brown}},\ }\href {\doibase 10.1088/0004-637X/722/1/33} {\bibfield  {journal}
  {\bibinfo  {journal} {Astrophys. J.}\ }\textbf {\bibinfo {volume} {722}},\
  \bibinfo {pages} {33} (\bibinfo {year} {2010})},\ \Eprint
  {http://arxiv.org/abs/1005.0811} {arXiv:1005.0811 [astro-ph.HE]} \BibitemShut
  {NoStop}%
%%CITATION = ARXIV:1005.0811;%%
\bibitem [{\citenamefont {Steiner}\ \emph {et~al.}(2013)\citenamefont
  {Steiner}, \citenamefont {Lattimer},\ and\ \citenamefont
  {Brown}}]{Steiner:2012xt}%
  \BibitemOpen
  \bibfield  {author} {\bibinfo {author} {\bibfnamefont {A.~W.}\ \bibnamefont
  {Steiner}}, \bibinfo {author} {\bibfnamefont {J.~M.}\ \bibnamefont
  {Lattimer}}, \ and\ \bibinfo {author} {\bibfnamefont {E.~F.}\ \bibnamefont
  {Brown}},\ }\href {\doibase 10.1088/2041-8205/765/1/L5} {\bibfield  {journal}
  {\bibinfo  {journal} {Astrophys. J.}\ }\textbf {\bibinfo {volume} {765}},\
  \bibinfo {pages} {L5} (\bibinfo {year} {2013})},\ \Eprint
  {http://arxiv.org/abs/1205.6871} {arXiv:1205.6871 [nucl-th]} \BibitemShut
  {NoStop}%
%%CITATION = ARXIV:1205.6871;%%
\bibitem [{\citenamefont {Raithel}\ \emph {et~al.}(2016)\citenamefont
  {Raithel}, \citenamefont {{\"O}zel},\ and\ \citenamefont
  {Psaltis}}]{Raithel:2016bux}%
  \BibitemOpen
  \bibfield  {author} {\bibinfo {author} {\bibfnamefont {C.~A.}\ \bibnamefont
  {Raithel}}, \bibinfo {author} {\bibfnamefont {F.}~\bibnamefont {{\"O}zel}}, \
  and\ \bibinfo {author} {\bibfnamefont {D.}~\bibnamefont {Psaltis}},\ }\href
  {\doibase 10.3847/0004-637X/831/1/44} {\bibfield  {journal} {\bibinfo
  {journal} {Astrophys. J.}\ }\textbf {\bibinfo {volume} {831}},\ \bibinfo
  {pages} {44} (\bibinfo {year} {2016})},\ \Eprint
  {http://arxiv.org/abs/1605.03591} {arXiv:1605.03591 [astro-ph.HE]}
  \BibitemShut {NoStop}%
%%CITATION = ARXIV:1605.03591;%%
\bibitem [{\citenamefont {Raithel}\ \emph {et~al.}(2017)\citenamefont
  {Raithel}, \citenamefont {{\"O}zel},\ and\ \citenamefont
  {Psaltis}}]{Raithel:2017ity}%
  \BibitemOpen
  \bibfield  {author} {\bibinfo {author} {\bibfnamefont {C.~A.}\ \bibnamefont
  {Raithel}}, \bibinfo {author} {\bibfnamefont {F.}~\bibnamefont {{\"O}zel}}, \
  and\ \bibinfo {author} {\bibfnamefont {D.}~\bibnamefont {Psaltis}},\ }\href
  {\doibase 10.3847/1538-4357/aa7a5a} {\bibfield  {journal} {\bibinfo
  {journal} {Astrophys. J.}\ }\textbf {\bibinfo {volume} {844}},\ \bibinfo
  {pages} {156} (\bibinfo {year} {2017})},\ \Eprint
  {http://arxiv.org/abs/1704.00737} {arXiv:1704.00737 [astro-ph.HE]}
  \BibitemShut {NoStop}%
%%CITATION = ARXIV:1704.00737;%%
\bibitem [{\citenamefont {Alvarez-Castillo}\ \emph {et~al.}(2016)\citenamefont
  {Alvarez-Castillo}, \citenamefont {Ayriyan}, \citenamefont {Benic},
  \citenamefont {Blaschke}, \citenamefont {Grigorian},\ and\ \citenamefont
  {Typel}}]{Alvarez-Castillo:2016oln}%
  \BibitemOpen
  \bibfield  {author} {\bibinfo {author} {\bibfnamefont {D.}~\bibnamefont
  {Alvarez-Castillo}}, \bibinfo {author} {\bibfnamefont {A.}~\bibnamefont
  {Ayriyan}}, \bibinfo {author} {\bibfnamefont {S.}~\bibnamefont {Benic}},
  \bibinfo {author} {\bibfnamefont {D.}~\bibnamefont {Blaschke}}, \bibinfo
  {author} {\bibfnamefont {H.}~\bibnamefont {Grigorian}}, \ and\ \bibinfo
  {author} {\bibfnamefont {S.}~\bibnamefont {Typel}},\ }\href {\doibase
  10.1140/epja/i2016-16069-2} {\bibfield  {journal} {\bibinfo  {journal} {Eur.
  Phys. J.}\ }\textbf {\bibinfo {volume} {A52}},\ \bibinfo {pages} {69}
  (\bibinfo {year} {2016})},\ \Eprint {http://arxiv.org/abs/1603.03457}
  {arXiv:1603.03457 [nucl-th]} \BibitemShut {NoStop}%
%%CITATION = ARXIV:1603.03457;%%
\bibitem [{\citenamefont {Ozel}\ \emph {et~al.}(2016)\citenamefont {Ozel},
  \citenamefont {Psaltis}, \citenamefont {Guver}, \citenamefont {Baym},
  \citenamefont {Heinke},\ and\ \citenamefont {Guillot}}]{Ozel:2015fia}%
  \BibitemOpen
  \bibfield  {author} {\bibinfo {author} {\bibfnamefont {F.}~\bibnamefont
  {Ozel}}, \bibinfo {author} {\bibfnamefont {D.}~\bibnamefont {Psaltis}},
  \bibinfo {author} {\bibfnamefont {T.}~\bibnamefont {Guver}}, \bibinfo
  {author} {\bibfnamefont {G.}~\bibnamefont {Baym}}, \bibinfo {author}
  {\bibfnamefont {C.}~\bibnamefont {Heinke}}, \ and\ \bibinfo {author}
  {\bibfnamefont {S.}~\bibnamefont {Guillot}},\ }\href {\doibase
  10.3847/0004-637X/820/1/28} {\bibfield  {journal} {\bibinfo  {journal}
  {Astrophys. J.}\ }\textbf {\bibinfo {volume} {820}},\ \bibinfo {pages} {28}
  (\bibinfo {year} {2016})},\ \Eprint {http://arxiv.org/abs/1505.05155}
  {arXiv:1505.05155 [astro-ph.HE]} \BibitemShut {NoStop}%
%%CITATION = ARXIV:1505.05155;%%
\bibitem [{\citenamefont {Bogdanov}\ \emph {et~al.}(2016)\citenamefont
  {Bogdanov}, \citenamefont {Heinke}, \citenamefont {{\"Ozel}},\ and\
  \citenamefont {{G\"uver}}}]{Bogdanov:2016nle}%
  \BibitemOpen
  \bibfield  {author} {\bibinfo {author} {\bibfnamefont {S.}~\bibnamefont
  {Bogdanov}}, \bibinfo {author} {\bibfnamefont {C.~O.}\ \bibnamefont
  {Heinke}}, \bibinfo {author} {\bibfnamefont {F.}~\bibnamefont {{\"Ozel}}}, \
  and\ \bibinfo {author} {\bibfnamefont {T.}~\bibnamefont {{G\"uver}}},\ }\href
  {\doibase 10.3847/0004-637X/831/2/184} {\bibfield  {journal} {\bibinfo
  {journal} {Astrophys. J.}\ }\textbf {\bibinfo {volume} {831}},\ \bibinfo
  {pages} {184} (\bibinfo {year} {2016})},\ \Eprint
  {http://arxiv.org/abs/1603.01630} {arXiv:1603.01630 [astro-ph.HE]}
  \BibitemShut {NoStop}%
%%CITATION = ARXIV:1603.01630;%%
\bibitem [{\citenamefont {Fujimoto}\ \emph {et~al.}(2018)\citenamefont
  {Fujimoto}, \citenamefont {Fukushima},\ and\ \citenamefont
  {Murase}}]{Fujimoto:2017cdo}%
  \BibitemOpen
  \bibfield  {author} {\bibinfo {author} {\bibfnamefont {Y.}~\bibnamefont
  {Fujimoto}}, \bibinfo {author} {\bibfnamefont {K.}~\bibnamefont {Fukushima}},
  \ and\ \bibinfo {author} {\bibfnamefont {K.}~\bibnamefont {Murase}},\ }\href
  {\doibase 10.1103/PhysRevD.98.023019} {\bibfield  {journal} {\bibinfo
  {journal} {Phys. Rev.}\ }\textbf {\bibinfo {volume} {D98}},\ \bibinfo {pages}
  {023019} (\bibinfo {year} {2018})},\ \Eprint
  {http://arxiv.org/abs/1711.06748} {arXiv:1711.06748 [nucl-th]} \BibitemShut
  {NoStop}%
%%CITATION = ARXIV:1711.06748;%%
\bibitem [{\citenamefont {Pang}\ \emph {et~al.}(2018)\citenamefont {Pang},
  \citenamefont {Zhou}, \citenamefont {Su}, \citenamefont {Petersen},
  \citenamefont {St{\"o}cker},\ and\ \citenamefont {Wang}}]{Pang:2016vdc}%
  \BibitemOpen
  \bibfield  {author} {\bibinfo {author} {\bibfnamefont {L.-G.}\ \bibnamefont
  {Pang}}, \bibinfo {author} {\bibfnamefont {K.}~\bibnamefont {Zhou}}, \bibinfo
  {author} {\bibfnamefont {N.}~\bibnamefont {Su}}, \bibinfo {author}
  {\bibfnamefont {H.}~\bibnamefont {Petersen}}, \bibinfo {author}
  {\bibfnamefont {H.}~\bibnamefont {St{\"o}cker}}, \ and\ \bibinfo {author}
  {\bibfnamefont {X.-N.}\ \bibnamefont {Wang}},\ }\href {\doibase
  10.1038/s41467-017-02726-3} {\bibfield  {journal} {\bibinfo  {journal}
  {Nature Commun.}\ }\textbf {\bibinfo {volume} {9}},\ \bibinfo {pages} {210}
  (\bibinfo {year} {2018})},\ \Eprint {http://arxiv.org/abs/1612.04262}
  {arXiv:1612.04262 [hep-ph]} \BibitemShut {NoStop}%
%%CITATION = ARXIV:1612.04262;%%
\bibitem [{\citenamefont {Mori}\ \emph {et~al.}(2018)\citenamefont {Mori},
  \citenamefont {Kashiwa},\ and\ \citenamefont {Ohnishi}}]{Mori:2017nwj}%
  \BibitemOpen
  \bibfield  {author} {\bibinfo {author} {\bibfnamefont {Y.}~\bibnamefont
  {Mori}}, \bibinfo {author} {\bibfnamefont {K.}~\bibnamefont {Kashiwa}}, \
  and\ \bibinfo {author} {\bibfnamefont {A.}~\bibnamefont {Ohnishi}},\ }\href
  {\doibase 10.1093/ptep/ptx191} {\bibfield  {journal} {\bibinfo  {journal}
  {PTEP}\ }\textbf {\bibinfo {volume} {2018}},\ \bibinfo {pages} {023B04}
  (\bibinfo {year} {2018})},\ \Eprint {http://arxiv.org/abs/1709.03208}
  {arXiv:1709.03208 [hep-lat]} \BibitemShut {NoStop}%
%%CITATION = ARXIV:1709.03208;%%
\bibitem [{\citenamefont {Niu}\ and\ \citenamefont
  {Liang}(2018)}]{Niu:2018csp}%
  \BibitemOpen
  \bibfield  {author} {\bibinfo {author} {\bibfnamefont {Z.~M.}\ \bibnamefont
  {Niu}}\ and\ \bibinfo {author} {\bibfnamefont {H.~Z.}\ \bibnamefont
  {Liang}},\ }\href {\doibase 10.1016/j.physletb.2018.01.002} {\bibfield
  {journal} {\bibinfo  {journal} {Phys. Lett.}\ }\textbf {\bibinfo {volume}
  {B778}},\ \bibinfo {pages} {48} (\bibinfo {year} {2018})},\ \Eprint
  {http://arxiv.org/abs/1801.04411} {arXiv:1801.04411 [nucl-th]} \BibitemShut
  {NoStop}%
%%CITATION = ARXIV:1801.04411;%%
\bibitem [{\citenamefont {George}\ and\ \citenamefont
  {Huerta}(2018)}]{George:2016hay}%
  \BibitemOpen
  \bibfield  {author} {\bibinfo {author} {\bibfnamefont {D.}~\bibnamefont
  {George}}\ and\ \bibinfo {author} {\bibfnamefont {E.~A.}\ \bibnamefont
  {Huerta}},\ }\href {\doibase 10.1103/PhysRevD.97.044039} {\bibfield
  {journal} {\bibinfo  {journal} {Phys. Rev.}\ }\textbf {\bibinfo {volume}
  {D97}},\ \bibinfo {pages} {044039} (\bibinfo {year} {2018})},\ \Eprint
  {http://arxiv.org/abs/1701.00008} {arXiv:1701.00008 [astro-ph.IM]}
  \BibitemShut {NoStop}%
%%CITATION = ARXIV:1701.00008;%%
\bibitem [{\citenamefont {Allen}\ \emph {et~al.}(2019)\citenamefont {Allen}
  \emph {et~al.}}]{Allen:2019dkq}%
  \BibitemOpen
  \bibfield  {author} {\bibinfo {author} {\bibfnamefont {G.}~\bibnamefont
  {Allen}} \emph {et~al.}\ }(\bibinfo {year} {2019})\ \Eprint
  {http://arxiv.org/abs/1902.00522} {arXiv:1902.00522 [astro-ph.IM]}
  \BibitemShut {NoStop}%
%%CITATION = ARXIV:1902.00522;%%
\bibitem [{\citenamefont {Douchin}\ and\ \citenamefont
  {Haensel}(2001)}]{Douchin:2001sv}%
  \BibitemOpen
  \bibfield  {author} {\bibinfo {author} {\bibfnamefont {F.}~\bibnamefont
  {Douchin}}\ and\ \bibinfo {author} {\bibfnamefont {P.}~\bibnamefont
  {Haensel}},\ }\href {\doibase 10.1051/0004-6361:20011402} {\bibfield
  {journal} {\bibinfo  {journal} {Astron. Astrophys.}\ }\textbf {\bibinfo
  {volume} {380}},\ \bibinfo {pages} {151} (\bibinfo {year} {2001})},\ \Eprint
  {http://arxiv.org/abs/astro-ph/0111092} {arXiv:astro-ph/0111092 [astro-ph]}
  \BibitemShut {NoStop}%
%%CITATION = ASTRO-PH/0111092;%%
\bibitem [{\citenamefont {Chollet}(2015)}]{software:Keras}%
  \BibitemOpen
  \bibfield  {author} {\bibinfo {author} {\bibfnamefont {F.}~\bibnamefont
  {Chollet}},\ }\href@noop {} {\enquote {\bibinfo {title} {Keras: Deep learning
  library for theano and tensorflow},}\ }\bibinfo {howpublished}
  {\url{https://github.com/fchollet/keras}} (\bibinfo {year}
  {2015})\BibitemShut {NoStop}%
\bibitem [{\citenamefont {Abadi}\ \emph {et~al.}(2016)\citenamefont {Abadi}
  \emph {et~al.}}]{arXiv:1605.08695}%
  \BibitemOpen
  \bibfield  {author} {\bibinfo {author} {\bibfnamefont {M.}~\bibnamefont
  {Abadi}} \emph {et~al.},\ }\href@noop {} {\  (\bibinfo {year} {2016})},\
  \Eprint {http://arxiv.org/abs/1605.08695} {arXiv:1605.08695 [cs.DC]}
  \BibitemShut {NoStop}%
\bibitem [{\citenamefont {LeCun}\ \emph {et~al.}(2015)\citenamefont {LeCun},
  \citenamefont {Bengio},\ and\ \citenamefont
  {Hinton}}]{DBLP:journals/nature/LeCunBH15}%
  \BibitemOpen
  \bibfield  {author} {\bibinfo {author} {\bibfnamefont {Y.}~\bibnamefont
  {LeCun}}, \bibinfo {author} {\bibfnamefont {Y.}~\bibnamefont {Bengio}}, \
  and\ \bibinfo {author} {\bibfnamefont {G.~E.}\ \bibnamefont {Hinton}},\
  }\href {\doibase 10.1038/nature14539} {\bibfield  {journal} {\bibinfo
  {journal} {Nature}\ }\textbf {\bibinfo {volume} {521}},\ \bibinfo {pages}
  {436} (\bibinfo {year} {2015})}\BibitemShut {NoStop}%
\bibitem [{\citenamefont {Kingma}\ and\ \citenamefont
  {Ba}(2014)}]{DBLP:journals/corr/KingmaB14}%
  \BibitemOpen
  \bibfield  {author} {\bibinfo {author} {\bibfnamefont {D.~P.}\ \bibnamefont
  {Kingma}}\ and\ \bibinfo {author} {\bibfnamefont {J.}~\bibnamefont {Ba}},\
  }\href {http://arxiv.org/abs/1412.6980} {\bibfield  {journal} {\bibinfo
  {journal} {CoRR}\ }\textbf {\bibinfo {volume} {abs/1412.6980}} (\bibinfo
  {year} {2014})},\ \Eprint {http://arxiv.org/abs/1412.6980} {arXiv:1412.6980}
  \BibitemShut {NoStop}%
\bibitem [{\citenamefont {Glorot}\ and\ \citenamefont
  {Bengio}(2010)}]{pmlr-v9-glorot10a}%
  \BibitemOpen
  \bibfield  {author} {\bibinfo {author} {\bibfnamefont {X.}~\bibnamefont
  {Glorot}}\ and\ \bibinfo {author} {\bibfnamefont {Y.}~\bibnamefont
  {Bengio}},\ }in\ \href {http://proceedings.mlr.press/v9/glorot10a.html}
  {\emph {\bibinfo {booktitle} {Proceedings of the Thirteenth International
  Conference on Artificial Intelligence and Statistics}}},\ \bibinfo {series}
  {Proceedings of Machine Learning Research}, Vol.~\bibinfo {volume} {9},\
  \bibinfo {editor} {edited by\ \bibinfo {editor} {\bibfnamefont {Y.~W.}\
  \bibnamefont {Teh}}\ and\ \bibinfo {editor} {\bibfnamefont {M.}~\bibnamefont
  {Titterington}}}\ (\bibinfo  {publisher} {PMLR},\ \bibinfo {address} {Chia
  Laguna Resort, Sardinia, Italy},\ \bibinfo {year} {2010})\ pp.\ \bibinfo
  {pages} {249--256}\BibitemShut {NoStop}%
\bibitem [{\citenamefont {Akmal}\ \emph {et~al.}(1998)\citenamefont {Akmal},
  \citenamefont {Pandharipande},\ and\ \citenamefont
  {Ravenhall}}]{Akmal:1998cf}%
  \BibitemOpen
  \bibfield  {author} {\bibinfo {author} {\bibfnamefont {A.}~\bibnamefont
  {Akmal}}, \bibinfo {author} {\bibfnamefont {V.~R.}\ \bibnamefont
  {Pandharipande}}, \ and\ \bibinfo {author} {\bibfnamefont {D.~G.}\
  \bibnamefont {Ravenhall}},\ }\href {\doibase 10.1103/PhysRevC.58.1804}
  {\bibfield  {journal} {\bibinfo  {journal} {Phys. Rev.}\ }\textbf {\bibinfo
  {volume} {C58}},\ \bibinfo {pages} {1804} (\bibinfo {year} {1998})},\ \Eprint
  {http://arxiv.org/abs/nucl-th/9804027} {arXiv:nucl-th/9804027 [nucl-th]}
  \BibitemShut {NoStop}%
%%CITATION = NUCL-TH/9804027;%%
\bibitem [{\citenamefont {Goriely}\ \emph {et~al.}(2010)\citenamefont
  {Goriely}, \citenamefont {Chamel},\ and\ \citenamefont
  {Pearson}}]{Goriely:2010bm}%
  \BibitemOpen
  \bibfield  {author} {\bibinfo {author} {\bibfnamefont {S.}~\bibnamefont
  {Goriely}}, \bibinfo {author} {\bibfnamefont {N.}~\bibnamefont {Chamel}}, \
  and\ \bibinfo {author} {\bibfnamefont {J.~M.}\ \bibnamefont {Pearson}},\
  }\href {\doibase 10.1103/PhysRevC.82.035804} {\bibfield  {journal} {\bibinfo
  {journal} {Phys. Rev.}\ }\textbf {\bibinfo {volume} {C82}},\ \bibinfo {pages}
  {035804} (\bibinfo {year} {2010})},\ \Eprint {http://arxiv.org/abs/1009.3840}
  {arXiv:1009.3840 [nucl-th]} \BibitemShut {NoStop}%
%%CITATION = ARXIV:1009.3840;%%
\bibitem [{\citenamefont {Engvik}\ \emph {et~al.}(1996)\citenamefont {Engvik},
  \citenamefont {Bao}, \citenamefont {Hjorth-Jensen}, \citenamefont {Osnes},\
  and\ \citenamefont {Ostgaard}}]{Engvik:1995gn}%
  \BibitemOpen
  \bibfield  {author} {\bibinfo {author} {\bibfnamefont {L.}~\bibnamefont
  {Engvik}}, \bibinfo {author} {\bibfnamefont {G.}~\bibnamefont {Bao}},
  \bibinfo {author} {\bibfnamefont {M.}~\bibnamefont {Hjorth-Jensen}}, \bibinfo
  {author} {\bibfnamefont {E.}~\bibnamefont {Osnes}}, \ and\ \bibinfo {author}
  {\bibfnamefont {E.}~\bibnamefont {Ostgaard}},\ }\href {\doibase
  10.1086/177827} {\bibfield  {journal} {\bibinfo  {journal} {Astrophys. J.}\
  }\textbf {\bibinfo {volume} {469}},\ \bibinfo {pages} {794} (\bibinfo {year}
  {1996})},\ \Eprint {http://arxiv.org/abs/nucl-th/9509016}
  {arXiv:nucl-th/9509016 [nucl-th]} \BibitemShut {NoStop}%
%%CITATION = NUCL-TH/9509016;%%
\bibitem [{\citenamefont {Hebeler}\ \emph {et~al.}(2013)\citenamefont
  {Hebeler}, \citenamefont {Lattimer}, \citenamefont {Pethick},\ and\
  \citenamefont {Schwenk}}]{Hebeler:2013nza}%
  \BibitemOpen
  \bibfield  {author} {\bibinfo {author} {\bibfnamefont {K.}~\bibnamefont
  {Hebeler}}, \bibinfo {author} {\bibfnamefont {J.~M.}\ \bibnamefont
  {Lattimer}}, \bibinfo {author} {\bibfnamefont {C.~J.}\ \bibnamefont
  {Pethick}}, \ and\ \bibinfo {author} {\bibfnamefont {A.}~\bibnamefont
  {Schwenk}},\ }\href {\doibase 10.1088/0004-637X/773/1/11} {\bibfield
  {journal} {\bibinfo  {journal} {Astrophys. J.}\ }\textbf {\bibinfo {volume}
  {773}},\ \bibinfo {pages} {11} (\bibinfo {year} {2013})},\ \Eprint
  {http://arxiv.org/abs/1303.4662} {arXiv:1303.4662 [astro-ph.SR]} \BibitemShut
  {NoStop}%
%%CITATION = ARXIV:1303.4662;%%
\bibitem [{\citenamefont {Mueller}\ and\ \citenamefont
  {Serot}(1996)}]{Mueller:1996pm}%
  \BibitemOpen
  \bibfield  {author} {\bibinfo {author} {\bibfnamefont {H.}~\bibnamefont
  {Mueller}}\ and\ \bibinfo {author} {\bibfnamefont {B.~D.}\ \bibnamefont
  {Serot}},\ }\href {\doibase 10.1016/0375-9474(96)00187-X} {\bibfield
  {journal} {\bibinfo  {journal} {Nucl. Phys.}\ }\textbf {\bibinfo {volume}
  {A606}},\ \bibinfo {pages} {508} (\bibinfo {year} {1996})},\ \Eprint
  {http://arxiv.org/abs/nucl-th/9603037} {arXiv:nucl-th/9603037 [nucl-th]}
  \BibitemShut {NoStop}%
%%CITATION = NUCL-TH/9603037;%%
\bibitem [{\citenamefont {Wiringa}\ \emph {et~al.}(1988)\citenamefont
  {Wiringa}, \citenamefont {Fiks},\ and\ \citenamefont
  {Fabrocini}}]{Wiringa:1988tp}%
  \BibitemOpen
  \bibfield  {author} {\bibinfo {author} {\bibfnamefont {R.~B.}\ \bibnamefont
  {Wiringa}}, \bibinfo {author} {\bibfnamefont {V.}~\bibnamefont {Fiks}}, \
  and\ \bibinfo {author} {\bibfnamefont {A.}~\bibnamefont {Fabrocini}},\ }\href
  {\doibase 10.1103/PhysRevC.38.1010} {\bibfield  {journal} {\bibinfo
  {journal} {Phys. Rev.}\ }\textbf {\bibinfo {volume} {C38}},\ \bibinfo {pages}
  {1010} (\bibinfo {year} {1988})}\BibitemShut {NoStop}%
%%CITATION = PHRVA,C38,1010;%%
\bibitem [{\citenamefont {Demorest}\ \emph {et~al.}(2010)\citenamefont
  {Demorest}, \citenamefont {Pennucci}, \citenamefont {Ransom}, \citenamefont
  {Roberts},\ and\ \citenamefont {Hessels}}]{Demorest:2010bx}%
  \BibitemOpen
  \bibfield  {author} {\bibinfo {author} {\bibfnamefont {P.}~\bibnamefont
  {Demorest}}, \bibinfo {author} {\bibfnamefont {T.}~\bibnamefont {Pennucci}},
  \bibinfo {author} {\bibfnamefont {S.}~\bibnamefont {Ransom}}, \bibinfo
  {author} {\bibfnamefont {M.}~\bibnamefont {Roberts}}, \ and\ \bibinfo
  {author} {\bibfnamefont {J.}~\bibnamefont {Hessels}},\ }\href {\doibase
  10.1038/nature09466} {\bibfield  {journal} {\bibinfo  {journal} {Nature}\
  }\textbf {\bibinfo {volume} {467}},\ \bibinfo {pages} {1081} (\bibinfo {year}
  {2010})},\ \Eprint {http://arxiv.org/abs/1010.5788} {arXiv:1010.5788
  [astro-ph.HE]} \BibitemShut {NoStop}%
%%CITATION = ARXIV:1010.5788;%%
\bibitem [{\citenamefont {Fonseca}\ \emph {et~al.}(2016)\citenamefont {Fonseca}
  \emph {et~al.}}]{Fonseca:2016tux}%
  \BibitemOpen
  \bibfield  {author} {\bibinfo {author} {\bibfnamefont {E.}~\bibnamefont
  {Fonseca}} \emph {et~al.},\ }\href {\doibase 10.3847/0004-637X/832/2/167}
  {\bibfield  {journal} {\bibinfo  {journal} {Astrophys. J.}\ }\textbf
  {\bibinfo {volume} {832}},\ \bibinfo {pages} {167} (\bibinfo {year}
  {2016})},\ \Eprint {http://arxiv.org/abs/1603.00545} {arXiv:1603.00545
  [astro-ph.HE]} \BibitemShut {NoStop}%
%%CITATION = ARXIV:1603.00545;%%
\bibitem [{\citenamefont {Antoniadis}\ \emph {et~al.}(2013)\citenamefont
  {Antoniadis} \emph {et~al.}}]{Antoniadis:2013pzd}%
  \BibitemOpen
  \bibfield  {author} {\bibinfo {author} {\bibfnamefont {J.}~\bibnamefont
  {Antoniadis}} \emph {et~al.},\ }\href {\doibase 10.1126/science.1233232}
  {\bibfield  {journal} {\bibinfo  {journal} {Science}\ }\textbf {\bibinfo
  {volume} {340}},\ \bibinfo {pages} {6131} (\bibinfo {year} {2013})},\ \Eprint
  {http://arxiv.org/abs/1304.6875} {arXiv:1304.6875 [astro-ph.HE]} \BibitemShut
  {NoStop}%
%%CITATION = ARXIV:1304.6875;%%
\bibitem [{\citenamefont {Cromartie}\ \emph {et~al.}(2019)\citenamefont
  {Cromartie} \emph {et~al.}}]{Cromartie:2019kug}%
  \BibitemOpen
  \bibfield  {author} {\bibinfo {author} {\bibfnamefont {H.~T.}\ \bibnamefont
  {Cromartie}} \emph {et~al.},\ }\href {\doibase 10.1038/s41550-019-0880-2} {\
  (\bibinfo {year} {2019}),\ 10.1038/s41550-019-0880-2},\ \Eprint
  {http://arxiv.org/abs/1904.06759} {arXiv:1904.06759 [astro-ph.HE]}
  \BibitemShut {NoStop}%
%%CITATION = ARXIV:1904.06759;%%
\bibitem [{\citenamefont {Gandolfi}\ \emph {et~al.}(2012)\citenamefont
  {Gandolfi}, \citenamefont {Carlson},\ and\ \citenamefont
  {Reddy}}]{Gandolfi:2011xu}%
  \BibitemOpen
  \bibfield  {author} {\bibinfo {author} {\bibfnamefont {S.}~\bibnamefont
  {Gandolfi}}, \bibinfo {author} {\bibfnamefont {J.}~\bibnamefont {Carlson}}, \
  and\ \bibinfo {author} {\bibfnamefont {S.}~\bibnamefont {Reddy}},\ }\href
  {\doibase 10.1103/PhysRevC.85.032801} {\bibfield  {journal} {\bibinfo
  {journal} {Phys. Rev.}\ }\textbf {\bibinfo {volume} {C85}},\ \bibinfo {pages}
  {032801} (\bibinfo {year} {2012})},\ \Eprint {http://arxiv.org/abs/1101.1921}
  {arXiv:1101.1921 [nucl-th]} \BibitemShut {NoStop}%
%%CITATION = ARXIV:1101.1921;%%
\bibitem [{\citenamefont {Bedaque}\ and\ \citenamefont
  {Steiner}(2015)}]{Bedaque:2014sqa}%
  \BibitemOpen
  \bibfield  {author} {\bibinfo {author} {\bibfnamefont {P.}~\bibnamefont
  {Bedaque}}\ and\ \bibinfo {author} {\bibfnamefont {A.~W.}\ \bibnamefont
  {Steiner}},\ }\href {\doibase 10.1103/PhysRevLett.114.031103} {\bibfield
  {journal} {\bibinfo  {journal} {Phys. Rev. Lett.}\ }\textbf {\bibinfo
  {volume} {114}},\ \bibinfo {pages} {031103} (\bibinfo {year} {2015})},\
  \Eprint {http://arxiv.org/abs/1408.5116} {arXiv:1408.5116 [nucl-th]}
  \BibitemShut {NoStop}%
%%CITATION = ARXIV:1408.5116;%%
\bibitem [{\citenamefont {Tews}\ \emph {et~al.}(2018)\citenamefont {Tews},
  \citenamefont {Carlson}, \citenamefont {Gandolfi},\ and\ \citenamefont
  {Reddy}}]{Tews:2018kmu}%
  \BibitemOpen
  \bibfield  {author} {\bibinfo {author} {\bibfnamefont {I.}~\bibnamefont
  {Tews}}, \bibinfo {author} {\bibfnamefont {J.}~\bibnamefont {Carlson}},
  \bibinfo {author} {\bibfnamefont {S.}~\bibnamefont {Gandolfi}}, \ and\
  \bibinfo {author} {\bibfnamefont {S.}~\bibnamefont {Reddy}},\ }\href
  {\doibase 10.3847/1538-4357/aac267} {\bibfield  {journal} {\bibinfo
  {journal} {Astrophys. J.}\ }\textbf {\bibinfo {volume} {860}},\ \bibinfo
  {pages} {149} (\bibinfo {year} {2018})},\ \Eprint
  {http://arxiv.org/abs/1801.01923} {arXiv:1801.01923 [nucl-th]} \BibitemShut
  {NoStop}%
%%CITATION = ARXIV:1801.01923;%%
\bibitem [{\citenamefont {Hinderer}\ \emph {et~al.}(2010)\citenamefont
  {Hinderer}, \citenamefont {Lackey}, \citenamefont {Lang},\ and\ \citenamefont
  {Read}}]{Hinderer:2009ca}%
  \BibitemOpen
  \bibfield  {author} {\bibinfo {author} {\bibfnamefont {T.}~\bibnamefont
  {Hinderer}}, \bibinfo {author} {\bibfnamefont {B.~D.}\ \bibnamefont
  {Lackey}}, \bibinfo {author} {\bibfnamefont {R.~N.}\ \bibnamefont {Lang}}, \
  and\ \bibinfo {author} {\bibfnamefont {J.~S.}\ \bibnamefont {Read}},\ }\href
  {\doibase 10.1103/PhysRevD.81.123016} {\bibfield  {journal} {\bibinfo
  {journal} {Phys. Rev.}\ }\textbf {\bibinfo {volume} {D81}},\ \bibinfo {pages}
  {123016} (\bibinfo {year} {2010})},\ \Eprint {http://arxiv.org/abs/0911.3535}
  {arXiv:0911.3535 [astro-ph.HE]} \BibitemShut {NoStop}%
%%CITATION = ARXIV:0911.3535;%%
\bibitem [{\citenamefont {Abbott}\ \emph {et~al.}(2018)\citenamefont {Abbott}
  \emph {et~al.}}]{Abbott:2018exr}%
  \BibitemOpen
  \bibfield  {author} {\bibinfo {author} {\bibfnamefont {B.~P.}\ \bibnamefont
  {Abbott}} \emph {et~al.} (\bibinfo {collaboration} {LIGO Scientific,
  Virgo}),\ }\href {\doibase 10.1103/PhysRevLett.121.161101} {\bibfield
  {journal} {\bibinfo  {journal} {Phys. Rev. Lett.}\ }\textbf {\bibinfo
  {volume} {121}},\ \bibinfo {pages} {161101} (\bibinfo {year} {2018})},\
  \Eprint {http://arxiv.org/abs/1805.11581} {arXiv:1805.11581 [gr-qc]}
  \BibitemShut {NoStop}%
%%CITATION = ARXIV:1805.11581;%%
\end{thebibliography}%

\end{document}